\documentclass[12pt]{article}

\usepackage{graphicx}
\usepackage{url}
\usepackage{amsfonts,amsmath,amssymb,amsthm}
\usepackage{fullpage}

\newcommand{\pone}{\mathcal{P}_1}
\newcommand{\ptwo}{\mathcal{P}_2}

\title{Discrete Voronoi Games and $\epsilon$-Nets,\\
in Two and Three Dimensions}

\author{Aritra Banik\thanks{Ben-Gurion University of the Negev} \and
		Jean-Lou De Carufel\thanks{Carleton University} \and
        Anil Maheshwari$^\dagger$\thanks{Research supported by NSERC.} \and
        Michiel Smid$^{\dagger\ddagger}$}

\begin{document}
\newtheorem{definition}{Definition}
\newtheorem{theorem}{Theorem}
\newtheorem{proposition}{Proposition}
\newtheorem{corollary}{Corollary}
\newtheorem{observation}{Observation}
\newtheorem{lemma}{Lemma}

\maketitle

\begin{abstract}
The one-round discrete Voronoi game, with respect to a $n$-point user set $U$,
consists of two players Player 1 ($\pone$) and Player 2 ($\ptwo$). At first, $\pone$ chooses a set of facilities $F_1$ following which $\ptwo$ chooses another set  of  facilities $F_2$, disjoint from $F_1$. The payoff of $\ptwo$ is defined as the cardinality of the set of points in $U$ which are closer to a facility in $F_2$ than to every facility in $F_1$, and the payoff of $\pone$ is the difference between the number of users in $U$ and the payoff of $\ptwo$. The objective of both the players in the game is to maximize their respective payoffs. In this paper we study the one-round discrete Voronoi game where $\pone$ places $k$ facilities and $\ptwo$ places one facility and we have denoted this game as $VG(k,1)$. Although the optimal solution of this game can be found in polynomial time, the polynomial has a very high degree. In this paper,
we focus on achieving approximate solutions to $VG(k,1)$
with significantly better running times. We provide a constant-factor approximate solution to the optimal strategy of $\pone$ in $VG(k,1)$ by establishing a connection between $VG(k,1)$ and weak $\epsilon$-nets.
To the best of our knowledge,
this is the first time that Voronoi games are studied from the point of view of $\epsilon$-nets.

\end{abstract}

\section{Introduction}\label{sec:intro}

Facility location is a sub-field of operations research and computational geometry that focuses on the optimal placement of facilities, subject to a set of constraints.
One of the most famous facility location problems
is the computation of the minimum enclosing disk.
Given a set $U$ of $n$ points in the plane,
compute the smallest circle that encloses $U$.
There is a $O(n)$ time algorithm by Megiddo~\cite{4568408} to solve this problem
as well as a $O(n)$ expected time algorithm by Welzl~\cite{Welzl91smallestenclosing}.
For an extensive discussion on geometric variants of the facility location problem,
refer to Fekete et al.~\cite{DBLP:conf/compgeom/FeketeMW00}.

\emph{Competitive} facility location
(or \emph{competitive spatial modelling})
is concerned
with the strategic placement of facilities by competing market players,
subject to a set of constraints.
In this setting,
each facility has its \emph{service zone},
consisting of the set of users it serves.
Different metrics can be used to determine the users served by a given facility.
In general,
the service zone of a player does not have to be connected.
In the continuous setting,
the objective of each player is to place a set of facilities
in order to maximize the total area of its service zone.
As for in the discrete setting,
the objective of each player is to place a set of facilities
in order to maximize the total number of users present in its service zone.
The study of competitive facility location
started by the work of Hotelling~\cite{hotelling1929} back in 1929.
The discrete setting was introduced by Eaton and Lipsey~\cite{eaton1975principle}.
Numerous variants of facility location problems have been studied
(refer to~\cite{eiselt_ejor,eiselt_ts,tobin} for comprehensive surveys).

\begin{figure}
\centering
\includegraphics[scale = 0.5]{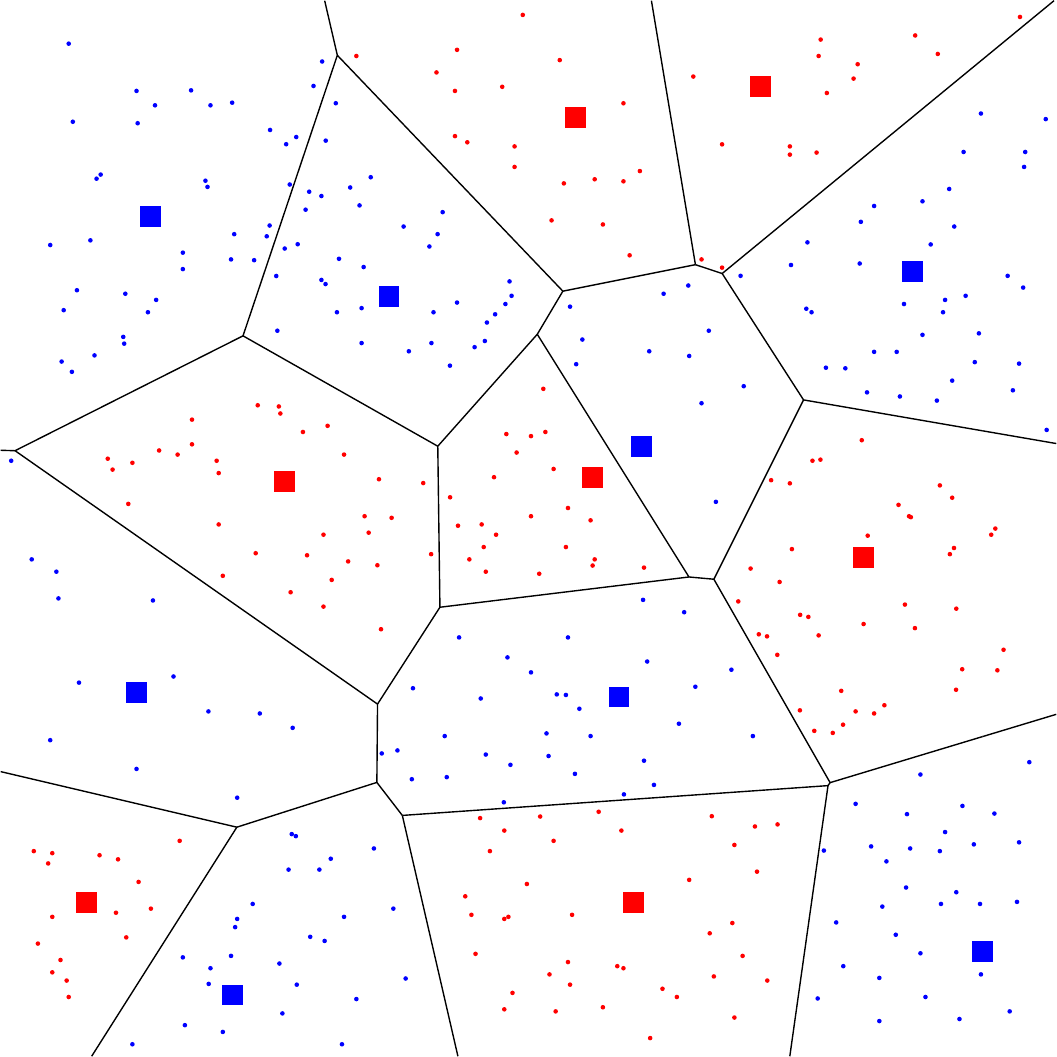}
\caption{A set $U$ of $n$ of users,
denoted by small points,
among two competing market players denoted by red and blue squares.
When we subdivide the space according to the nearest-neighbour rule,
we get the Voronoi diagram of the set of facilities.
\label{fig:distofusers}}
\end{figure}

\emph{Voronoi games},
introduced by Ahn et al.~\cite{DBLP:conf/cocoon/AhnCCGO01}
(for the one-dimensional case),
and by Cheong et al.~\cite{Cheong:2002:OVG:513400.513413}
(for two- and higher-dimensional cases),
consist in the following competitive facility location problem.
Two players alternately place one facility in $\mathbb{R}^d$,
until each of them has placed a given number of points.
Then we subdivide the space according to the nearest-neighbour rule.
The player whose facilities control the larger volume wins.
In the discrete setting,
introduced by Banik et al.~\cite{banik,banik2011},
the players are given a set $U$ of $n$ users in $\mathbb{R}^d$
(refer to Figure~\ref{fig:distofusers}).
Then,
as in the continuous setting,
the players alternately place one facility,
until each of them has placed a given number of points.
Then we subdivide the space according to the nearest-neighbour rule.
The player whose service zone contains the largest number of users wins.
To \emph{solve} a Voronoi game corresponds
to finding an optimal strategy for each player.
In this paper,
we establish a connection between Voronoi games and weak $\varepsilon$-nets,
which leads to general bounds on the scores of the players
for discrete Voronoi games.

\subsection{Preliminaries and Related Work}
\label{subsection preliminaries}

In this section,
we introduce the notation we use throughout the paper,
we define the variants of Voronoi games we study
and we explain how these variants compare to existing ones.
Consider a set $U$ of $n$ users
and let $F$ be a finite set of facilities.
In this paper,
we identify each facility with its location.
For every facility $f\in F$,
we define the \emph{service zone of $f$},
noted $U(f,F)$,
as the set of users from $U$ that are served by $f$.
The \emph{discrete Voronoi game} is a competitive facility location problem
involving two players $\pone$ and $\ptwo$,
respectively.
The players $\pone$ and $\ptwo$ alternately place
two disjoint sets of facilities
$F_1$ and $F_2$,
respectively.
A user $u\in U$ is said to be served by $\pone$
(respectively by $\ptwo$)
if $u$ belongs to the service zone of at least one facility placed by $\pone$
(respectively by $\ptwo$).
In such a case,
we say that $u$ is in the service zone of $\pone$
(respectively of $\ptwo$).
The \emph{payoff $\mathcal{V}(F_1, F_2)$ of $\ptwo$} (or the value of the game) is defined as the cardinality of the set of users from $U$ that belong to the service zone of $\ptwo$.
More formally,
$$\mathcal{V}(F_1, F_2)=\left|\bigcup_{f\in F_2}U(f, F_1\cup F_2)\right| .$$
Similarly,
the payoff of $\pone$ is
$$|U|- \mathcal{V}(F_1, F_2)=\left|\bigcup_{f\in F_1}U(f, F_1\cup F_2)\right| .$$
If $\mathcal{V}(F_1, F_2) > |U|- \mathcal{V}(F_1, F_2)$,
we say that $\ptwo$ wins.
If $\mathcal{V}(F_1, F_2) < |U|- \mathcal{V}(F_1, F_2)$,
we say that $\pone$ wins.
Otherwise,
it is declared a tie.
In the \emph{one-round Voronoi Game},
$\pone$ places all its facilities after which
$\ptwo$ places all its facilities.
If $|F_1| = |F_2| = k$,
the \emph{$k$-round Voronoi Game}
corresponds to a Voronoi game where the players
alternately place their facilities one at a time
(refer to~\cite{cccg2013}).

Let us define the \emph{One-Round Discrete Voronoi Game}.
\begin{definition}[One-Round Discrete Voronoi Game ${VG}(k,l)$]
Let $U$ be a set of $n$ users and $\pone$ and $\ptwo$ be two players. 
Initially, $\pone$ chooses a set $F_1$ of $k$ locations in $\mathbb{R}^d$ for its facilities.
Then $\ptwo$ chooses a set $F_2$ of $l$ locations in $\mathbb{R}^d$ for its facilities,
where $F_1\cap F_2={\O}$.
\begin{enumerate}
\item Given any choice of $F_1$ by $\pone$,
the objective of $\ptwo$ is to find a set
$F_2^*$ of $l$ points,
disjoint from $F_1$, that maximizes 
$\mathcal{V}(F_1,F_2)$,
where the maximum is taken over all sets of $l$ points $F_2\subset\mathbb{R}^d$
with $F_1\cap F_2={\O}$.
Formally,
the objective of $\ptwo$ is to find a set $F_2^*$ of $l$ points such that
$$\mathcal{V}(F_1,F_2^*) = \max_{\substack{F_2\subset\mathbb{R}^d\\ F_1\cap F_2={\O}\\ |F_2|=l}}\mathcal{V}(F_1,F_2) .$$

\item The objective of $\pone$ is to choose a set $F_1^*$ of 
$k$ facilities to minimize the maximum payoff of $\ptwo$.
In other words,
the objective of $\pone$ is to find a set $F_1^*$ of $k$ points such that
$$\max_{\substack{F_2\subset\mathbb{R}^d\\ F_1\cap F_2={\O}\\ |F_2|=l}}\mathcal{V}(F_1,F_2) $$
is minimized when $F_1 = F_1^*$,
where the minimum is taken of over all sets of $k$ points $F_1\subset\mathbb{R}^d$
with $F_1\cap F_2={\O}$.
\end{enumerate}
This game is called the \emph{One-Round Discrete Voronoi Game},
noted $VG(k,l)$.
\end{definition}

In this paper,
we study $VG(k,1)$ for all $k\geq 1$,
in two and three dimensions.

Voronoi games were introduced\footnote{Notice
that Shasha~\cite{shasha},
in 1992,
described a game called the \emph{Territory game}.
which corresponds to a variant of a discrete Voronoi game.} by Ahn et al.~\cite{DBLP:conf/cocoon/AhnCCGO01}
(for the one-dimensional case),
and by Cheong et al.~\cite{Cheong:2002:OVG:513400.513413}
(for two- and higher-dimensional cases).
The discrete version was introduced by Banik et al.~\cite{banik,banik2011}.
They first studied a version of ${VG}(k,k)$,
where the users in $U$ are restricted to a line.
They showed that,
if the sorted order of points in $U$ along the line is given,
then for any given placement of facilities of $\pone$,
an optimal strategy for $\ptwo$ can be computed in linear time.
They also provided results for determining an optimal strategy for $\pone$.
Then,
Banik et al.~\cite{cccg2013}
studied the case where $U$ can be any finite subset of $\mathbb{R}^2$.
They focused on the following version of the game.
The players $\pone$ and $\ptwo$ already own two sets of facilities $F_1$ and $F_2$,
respectively. The player $\pone$ wants to place one more facility knowing that $\ptwo$ will place another facility afterwards.
This game is called the
\emph{One-Round Discrete Voronoi Game in Presence of Existing Facilities},
or $VG_{1}^{\exists}(F_1,F_2)$ for short.
The optimal strategy of $\ptwo$, given any placement of $\pone$, is identical to the solution of the \emph{MaxCov} problem studied by Cabello et al.~\cite{reverse}. Consider a set $U$ of users, two sets of facilities $F_1$ and $F_2$, and any placement of a new facility $f$ by $\pone$. Let $U_1\subseteq U$ denote the subset of users that are served by $\pone$, in presence of $F_1$, $F_2$, and $f$. For every point $u\in U_1$, consider the \emph{nearest facility disk} $C_u$ centered at $u$ and passing through the facility in $F_1\cup \{f\}$ which is closest to $u$. Note that a new facility $f'$ placed by $\ptwo$ serves any user $u \in U_1$ if and only if $f'\in C_u$. Let $\mathcal C=\{C_u| u\in U_1\}$. Any optimal strategy for $\ptwo$ is a point which is inside a maximum number of circles among the circles in $\mathcal C$. This is the problem of finding the maximum depth in an arrangement of $n$ disks, and can be computed in $O(n^2)$ time~\cite{depthdisk}. Banik et al.~\cite{cccg2013} study how this arrangement 
changes as $f$ and $f'$ move in the plane. They provide a complete characterization of the event points and obtain an algorithm running in $O(n^8)$ time for computing an optimal placement of $P_1$.

Chawla et al.~\cite{Chawla:2004:WPL:988772.988815,DBLP:journals/orl/ChawlaRRS06} studied the impact
of knowing versus not knowing
the number of rounds in the game.
Suppose that $\pone$ does not know in advance the number of rounds,
whereas $\ptwo$ does.
What is the impact on the payoffs of $\pone$ and $\ptwo$?
Chawla et al. describe how $\pone$ suffers from this lack of information.
They worked in a framework where the distribution of the users is continuous,
but the set of possible locations for the players is finite.
They provide bounds on the payoff of $\pone$ in that framework.

The game $VG_{1}^{\exists}(F_1,F_2)$ where $F_2 = {\O}$
resembles $VG(k,1)$.
The difference is that in $VG(k,1)$,
$\pone$ can choose the location of all of its facilities
whereas in $VG_{1}^{\exists}(F_1,{\O})$,
$\pone$ can only choose the location of one facility.
Moreover,
the solution to $VG_{1}^{\exists}(F_1,{\O})$,
takes polynomial time,
where the polynomial has a very high degree.
In this paper,
we focus on achieving approximate solutions to $VG(k,1)$
with significantly better running times.

We provide a constant-factor approximate solution to the optimal strategy of $\pone$ in $VG(k,1)$.
We establish a connection between $VG(k,1)$ and weak $\epsilon$-nets.
To the best of our knowledge,
this is the first time that Voronoi game is studied from the point of view of $\epsilon$-nets.
We propose two different approaches that lead to different bounds
(refer to Theorems~\ref{theorem approx general k}
and~\ref{theorem approx general k 7 points}).
We give a bound on the values of $k$ for which
the first approach is better than the second one
(refer to Proposition~\ref{proposition approach 1 better}).
As we present the different results,
we discuss the algorithmic aspects of the computation of these approximate solutions.
We prove that for $k\geq 5$,
even though our strategy is approximate,
it guarantees that $\pone$ wins $VG(k,1)$
(refer to Corollary~\ref{corollary pone wins 2D}).
These results are presented in Section~\ref{section 2D}.
The first approach is the subject of Subsection~\ref{section k <= 136}
and the second approach is presented in Subsection~\ref{section k >= 137}.
Analogous results for $VG(k,1)$ in three dimensions
are presented in Section~\ref{section 3D}.
We discuss several open problems in Section~\ref{section conclusion}.

\section{Approximate Solutions for $VG(k,1)$}
\label{section 2D}

In this section,
we establish a connection between $VG(k,l)$ and weak $\epsilon$-nets.
This connection leads to an approximate solution and general bounds
on the payoff of $\pone$ in $VG(k,1)$.
For the rest of this section,
$U$ denotes a set of $n$ users in the plane,
in general position.
We study the specific cases of $VG(1,1)$
in Subsection~\ref{subsection label VG(1,1) 2D}
and $VG(2,1)$ in Subsection~\ref{subsection label VG(2,1) 2D}.
For the general case of $VG(k,1)$ ($k \geq 1$),
we propose two different approaches
(both related to weak $\epsilon$-nets).
We present them in Subsections~\ref{section k <= 136}
and~\ref{section k >= 137},
respectively,
and compare them in Subsection~\ref{section k >= 137}.

\subsection{An Approximate Solution for $VG(1,1)$}
\label{subsection label VG(1,1) 2D}

Using an approach similar to that of Banik et al.~\cite{cccg2013},
we can solve $VG(1,1)$ exactly.
However,
we get a polynomial time algorithm for finding the optimal strategy for $\pone$,
where the polynomial has a high degree.
This is why in this subsection,
as well as in the rest of the paper,
we turn to approximation algorithms.
The results in this subsection
are subsumed in~\cite{DBLP:journals/orl/ChawlaRRS06}.
Nonetheless,
we provide proofs for all results since they help understanding
the subsequent proofs in the paper.

We start with the following lemma.
\begin{lemma}
\label{lemma upper bound k = 1}
Let $f_1 \in \mathbb{R}^2$ be the location of the facility placed by $\pone$.
There is a strategy for $\ptwo$ that guarantees a payoff of at least $\frac{1}{2}n$ for $\ptwo$.
\end{lemma}

\proof
Let $\ell$ be any line passing through $f_1$.
The line $\ell$ defines two half-planes $H_1$ and $H_1'$, respectively.
Without loss of generality,
$H_1$ contains at least half of the users from $U$.
Therefore,
by placing one facility $f'\in H_1$ arbitrarily close to $f_1$,
$\ptwo$ can serve at least $\frac{1}{2}n$ users.
\qed

In the proof of Lemma~\ref{lemma upper bound k = 1},
we ensure that $\ptwo$ serves at least $\frac{1}{2}n$ users
by placing its facility arbitrarily close to $f_1$.
The argument works because of the general position assumption.
The same idea is used at several places in the paper.

A direct consequence of Lemma~\ref{lemma upper bound k = 1}
is that the payoff of $\pone$ is at most $\frac{1}{2}n$.
To find a lower bound on the payoff of $\pone$,
we consider the \emph{centerpoint}
(refer to~\cite[p.14]{Matousek:2002:LDG:581165})
of $U$.
\begin{definition}[Centerpoint]
Let $P$ be a set of $n$ points in $\mathbb{R}^2$.
A point $x\in \mathbb{R}^2$ is called a \emph{centerpoint} of $P$
if each closed half-space containing $x$
contains at least $\frac{1}{3}n$ points of $P$.
\end{definition}
We have the following theorem
about centerpoints
(refer to~\cite[Centerpoint theorem
]{Matousek:2002:LDG:581165}).
\begin{theorem}
\label{theorem centerpoint}
Each finite set of points in $\mathbb{R}^2$ has at least one centerpoint.
\end{theorem}

With Theorem~\ref{theorem centerpoint},
we can prove the following lemma.
\begin{lemma}
If $\pone$ places its facilities at the centerpoint of $U$,
then the payoff of $\ptwo$ is at most $\frac{2}{3}n$.
\end{lemma}

\proof
Let $f'$ be the location of the facility of $\ptwo$.
Player $\ptwo$ serves the set of users present in the Voronoi region of $f'$
in the Voronoi diagram of $f_1$ and $f'$.
However,
the Voronoi region of $f_1$ is a half-plane $H_1$ which contains $f_1$.
Therefore,
since $f_1$ is the centerpoint of $U$,
$H_1$ contains at least $\frac{1}{3}n$ users.
Thus,
$\ptwo$ can serve at most $\frac{2}{3}n$ users.
\qed

Therefore,
$$\frac{1}{3}n \leq \text{payoff of $\pone$} \leq \frac{1}{2}n .$$
Consequently,
we get a $\frac{n/2}{n/3} = \frac{3}{2}$-factor approximation of the optimal strategy for $\pone$.
Jadhav and Mukhopadhyay~\cite{Mukhopadhyay1994} showed how to compute the centerpoint of a set of $n$ points in $O(n)$ time,
which leads to the following theorem.
\begin{theorem}
We can compute in $O(n)$ time a $\frac{3}{2}$-factor approximation of the optimal strategy for $\pone$
in $VG(1,1)$.
\end{theorem}

\subsection{An Approximate Solution for $VG(2,1)$}
\label{subsection label VG(2,1) 2D}

In this subsection,
we present an approximate solution to $VG(2,1)$.
We start with the following lemma.
\begin{lemma}
\label{lemma upper bound k = 2}
Let $f_1,f_2 \in \mathbb{R}^2$ be the locations of the two facilities placed by $\pone$.
There is a strategy for $\ptwo$ that guarantees a payoff of at least $\frac{1}{4}n$ for $\ptwo$.
\end{lemma}

\begin{figure}
\centering
\includegraphics[scale = 1]{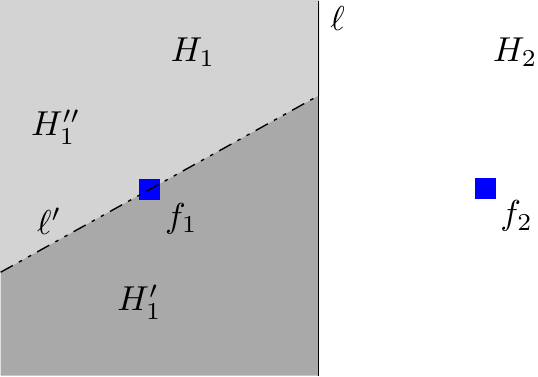}
\caption{Illustration of the proof of Lemma~\ref{lemma upper bound k = 2}.\label{fig:n/4}}
\end{figure}

\proof
Let $\ell$ be the perpendicular bisector of $f_1$ and $f_2$.
The line $\ell$ splits the plane into two half-planes
$H_1$ and $H_2$ respectively
containing $f_1$ and $f_2$
(refer to Figure~\ref{fig:n/4}).
Without loss of generality,
suppose that $H_1$ contains at least $\frac{1}{2}n$ users.

Let $\ell'$ be any line passing through $f_1$.
The line $\ell'$ defines two half-planes $H_1'$ and $H_1''$, respectively.
Without loss of generality,
$H_1'$ contains at least half of the users from $H_1$.
Therefore,
by placing one facility $f'\in H_1'$ arbitrarily close to $f_1$,
$\ptwo$ can serve at least $\frac{1}{4}n$ users.
\qed

A direct consequence of Lemma~\ref{lemma upper bound k = 2}
is that the payoff of $\pone$ is at most $\frac{3}{4}n$.
To find a lower bound on the payoff of $\pone$,
we use weak $\epsilon$-nets
(refer to~\cite{Matousek:2002:LDG:581165}).
\begin{definition}[Weak $\epsilon$-net]
Consider any real number $\epsilon\in[0,1]$. Let $X$ be a finite set of points in $\mathbb{R}^2$ and $\mathcal{R}$ be a set of subsets of $X$. We call the pair $(X, \mathcal{R})$ a \emph{range space}. The elements of $X$ and $\mathcal{R}$ are called \emph{points} and \emph{ranges} of the range space, respectively.

Let $N\subseteq \mathbb{R}^2$ be a finite set
such that $N$ intersects every set $K\in \mathcal{R}$ with $|K|>\epsilon|X|$.
If $N\subseteq X$,
then $N$ is an \emph{$\epsilon$-net} for $(X, \mathcal{R})$.
Otherwise,
$N$ is a \emph{weak $\epsilon$-net} for $(X, \mathcal{R})$.
\end{definition}

Mustafa and Ray~\cite{nabilsaurabh} proved the following theorem.
\begin{theorem}[Mustafa and Ray (2009)]
\label{th:rayndmustafa}
Let $P$ be a set of $n$ points in $\mathbb{R}^2$.
There exist two distinct points $z_1(P)$ and $z_2(P)$
such that any convex set containing more than $\frac{4}{7}n$ points of $P$
also contains at least one point from $\{z_1(P),z_2(P)\}$.
\end{theorem}

With Theorem~\ref{th:rayndmustafa},
we can prove the following lemma.
\begin{lemma}
If $\pone$ places its facilities at $z_1(U)$ and $z_2(U)$, respectively,
then the payoff of $\ptwo$ is at most $\frac{4}{7}n$.
\end{lemma}

\proof
Suppose that $f_1 = z_1(U)$, $f_2 = z_2(U)$
and $\ptwo$ places its facility at a point $f'$ such that its payoff is more than $\frac{4}{7}n$.
Player $\ptwo$ serves the set of users present in the Voronoi region of $f'$ in the Voronoi diagram of $f_1,f_2$ and $f'$. However,
the Voronoi region of $f'$ is a convex set which does not contain any of $f_1$ and $f_2$ but contains more than $\frac{4}{7}n$ points of $U$. This contradicts Theorem~\ref{th:rayndmustafa}.
\qed

Therefore,
$$\frac{3}{7}n \leq \text{payoff of $\pone$} \leq \frac{3}{4}n .$$
Consequently,
we get a $\frac{3n/4}{3n/7} = \frac{7}{4}$-factor approximation of the optimal strategy for $\pone$.
Langerman et al.~\cite{langerman2012computing} showed how to compute $z_1(U)$ and $z_2(U)$
in $O(n\log^4(n))$ time,
which leads to the following theorem.
\begin{theorem}
We can compute in $O(n\log^4(n))$ time a $\frac{7}{4}$-factor approximation of the optimal strategy for $\pone$
in $VG(2,1)$.
\end{theorem}

If we look into the proof of Theorem~\ref{th:rayndmustafa},
we see that one of the points in $\{z_1(P),z_2(P)\}$
is a point from $P$,
but in general,
the other point is not from $P$.
So in general,
$\{z_1(P),z_2(P)\}$ is a weak $\epsilon$-net.

\subsection{Bounds for $VG(k,1)$, Where $k \leq 136$ }
\label{section k <= 136}

In this section,
we study the more general case of $VG(k,1)$ for $k \geq 2$.
We start with the following lemma.
\begin{lemma}
\label{lemma general upper bound}
Let $\{f_1,f_2,...,f_k\} \subset \mathbb{R}^2$ be the set of locations of the $k$ facilities placed by $\pone$.
There is a strategy for $\ptwo$ that guarantees a payoff of at least $\frac{1}{2k}n$ for $\ptwo$.
\end{lemma}

\begin{figure}
\centering
\includegraphics[scale = 1]{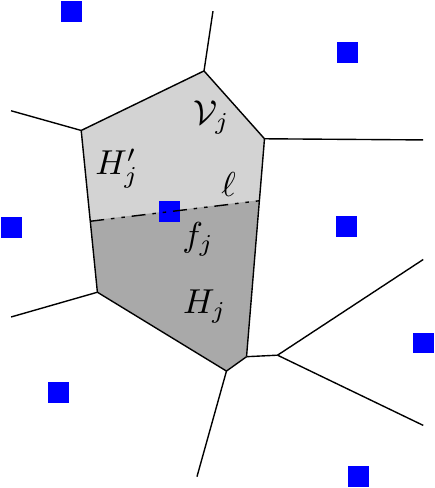}
\caption{Illustration of the proof of Lemma~\ref{lemma general upper bound}.\label{fig:n/k}}
\end{figure}

\proof
Consider the Voronoi diagram of $\{f_1,f_2,...,f_k\}$.
It consists in $k$ Voronoi cells
(refer to Figure~\ref{fig:n/k}).
Without loss of generality,
suppose that $\mathcal{V}_j$, the Voronoi cell of $f_j$,
contains at least $\frac{1}{k}n$ users.

Let $\ell$ be any line passing through $f_j$.
The line $\ell$ defines two half-planes $H_j$ and $H_j'$, respectively.
Without loss of generality,
$H_j$ contains at least half of the users in $\mathcal{V}_j$.
Therefore,
by placing one facility $f'\in H_j$ arbitrarily close to $f_j$,
$\ptwo$ can serve at least $\frac{1}{2k}n$ users.
\qed

A direct consequence of Lemma~\ref{lemma general upper bound}
is that the payoff of $\pone$ is at most $\frac{2k-1}{2k}n$.
To find a lower bound on the payoff of $\pone$,
we use a theorem by Mustafa and Ray~\cite{nabilsaurabh}.
Let us first introduce some notation.
Let $P$ be a set of $n$ points in $\mathbb{R}^d$.
Denote by $\epsilon_i^d$ the smallest number such that
there exists a set $N$ of $i$ distinct points satisfying the following property:
any convex set containing more than $\epsilon_i^dn$ points of $P$
also contains at least one point of $N$.

Given a set $\mathcal{K}$ of convex sets,
we say that $P$ \emph{pierces} $\mathcal{K}$ if for any disk $K\in \mathcal{K}$,
$K\cap P\neq {\O}$.
To the best of our knowledge,
the only exact values of $\epsilon_i^d$'s that are known are
$\epsilon_1^d = \frac{d}{d+1}$
and $\epsilon_2^2=\frac{4}{7}$
(refer to~\cite[Proposition 3.1]{nabilsaurabh}).
They are optimal in the following sense.
There exist arbitrarily large point sets such that
the set of all convex sets containing $\epsilon_i^ d n$ points
cannot be pierced by $i$ points.

Mustafa and Ray~\cite[Theorem 2.1]{nabilsaurabh} proved the following inequality.
\begin{theorem}[Mustafa and Ray (2009)]
Given any integers $r \geq 0$ and $s \geq 0$,
we have
\begin{align}
\label{ineq epsilon i d}
\epsilon_{r+ds+1}^d \leq \frac{\epsilon_r^d\left(1+(d-1)\epsilon_s^d\right)}{1+\epsilon_r^d\left(1+(d-1)\epsilon_s^d\right)},
\end{align}
where $\epsilon_0^d = 1$.
\end{theorem}
Even though~\eqref{ineq epsilon i d} does not lead
to the exact values of $\epsilon_i^d$'s in general,
we can still use it to find upper bounds on the $\epsilon_i^d$'s.
For all integers $d \geq 2$ and $i \geq 0$,
we define
\begin{align}
\label{definition overline epsilon}
\overline{\epsilon_i^d} = \begin{cases}
1 & \text{if } i = 0 , \cr
\min\limits_{\substack{r,s\geq 0\\r+ds+1=i}}\frac{\overline{\epsilon_r^d}\left(1+(d-1)\overline{\epsilon_s^d}\right)}{1+\overline{\epsilon_r^d}\left(1+(d-1)\overline{\epsilon_s^d}\right)} & i > 0 .\cr
\end{cases}
\end{align}
For all $d \geq 2$ and $i \geq 0$,
we have $\epsilon_i^d \leq \overline{\epsilon_i^d}$.
We can prove it by induction using the fact
that if $\epsilon_r^d \leq \overline{\epsilon_r^d}$
and $\epsilon_s^d \leq \overline{\epsilon_s^d}$,
then
\begin{align*}
\epsilon_i^d
&\leq \min\limits_{\substack{r,s\geq 0\\r+ds+1=i}}\frac{\epsilon_r^d\left(1+(d-1)\epsilon_s^d\right)}{1+\epsilon_r^d\left(1+(d-1)\epsilon_s^d\right)} & \text{by~\eqref{ineq epsilon i d},}\\
&= \min\limits_{\substack{r,s\geq 0\\r+ds+1=i}}\frac{1}{\frac{1}{\epsilon_r^d\left(1+(d-1)\epsilon_s^d\right)}+1} \\
&\leq \min\limits_{\substack{r,s\geq 0\\r+ds+1=i}}\frac{1}{\frac{1}{\overline{\epsilon_r^d}\left(1+(d-1)\overline{\epsilon_s^d}\right)}+1} \\
&= \min\limits_{\substack{r,s\geq 0\\r+ds+1=i}}\frac{\overline{\epsilon_r^d}\left(1+(d-1)\overline{\epsilon_s^d}\right)}{1+\overline{\epsilon_r^d}\left(1+(d-1)\overline{\epsilon_s^d}\right)} \\
&=\overline{\epsilon_i^d} & \text{by~\eqref{definition overline epsilon}.}
\end{align*}
Using a similar argument,
we can show that for a fixed $d$,
the sequence $\left(\overline{\epsilon_i^d}\right)_{i\geq 0}$ is decreasing.

To prove their theorem
(refer to~\eqref{ineq epsilon i d}),
Mustafa and Ray~\cite{nabilsaurabh} show the following.
\begin{corollary}[Mustafa and Ray (2009)]
\label{cor:rayndmustafa general}
Let $P$ be a set of $n$ points in $\mathbb{R}^d$.
There exists a set $E_i^d=\{z_1(P),z_2(P),...,z_i(P)\}$ of $i$ distinct points
such that any convex set containing more than $\overline{\epsilon_i^d}n$ points of $P$
also contains at least one point of $E_i^d$.
\end{corollary}

We can now give an upper bound on the payoff of $\ptwo$.
\begin{lemma}
\label{lemma ptwo at most epsilon k 2}
If $\pone$ places its facilities at all points of $E_k^2$, respectively,
then the payoff of $\ptwo$ is at most $\overline{\epsilon_k^2}n$.
\end{lemma}

\proof
Suppose that $f_i = z_i(U)$ ($1 \leq i \leq k$)
and $\ptwo$ places its facility at a point $f'$ such that its payoff
is more than $\overline{\epsilon_i^d}n$.
Player $\ptwo$ serves the set of users present in the Voronoi region of $f'$ in the Voronoi diagram of $E_k^2 \cup \{f'\}$. However,
the Voronoi region of $f'$ is a convex set which does not contain any point from $E_k^2$ but contains more than $\overline{\epsilon_i^d}n$ points of $U$. This contradicts Corollary~\ref{cor:rayndmustafa general}.
\qed

Therefore,
\begin{align}
\label{bounds pone < 136}
\left(1-\overline{\epsilon_k^2}\right)n \leq \text{payoff of $\pone$} \leq \frac{2k-1}{2k}n .
\end{align}
Consequently,
we get a $\frac{(2k-1)n/2k}{(1-\overline{\epsilon_k^2})n} = \frac{2k-1}{2k(1-\overline{\epsilon_k^2})}$-factor approximation of the optimal strategy for $\pone$.
This leads to the following theorem.
\begin{theorem}
\label{theorem approx general k}
We can compute in $O(kn\log^4(n))$ time a $\frac{2k-1}{2k\left(1-\overline{\epsilon_k^2}\right)}$-factor approximation of the optimal strategy for $\pone$
in $VG(k,1)$.
\end{theorem}

\proof
We already explained how,
by computing $E_k^2$,
we can get a $\frac{2k-1}{2k\left(1-\overline{\epsilon_k^2}\right)}$-factor approximation of the optimal strategy for $\pone$
in $VG(k,1)$.
It remains to explain how to compute $E_k^2$ in $O(kn\log^4(n))$ time.

Using dynamic programming,
we can compute and store the numbers $r_i$ and $s_i$
($0 \leq i \leq k$) such that
$$\overline{\epsilon_i^d} = \frac{\overline{\epsilon_{r_i}^d}\left(1+(d-1)\overline{\epsilon_{s_i}^d}\right)}{1+\overline{\epsilon_{r_i}^d}\left(1+(d-1)\overline{\epsilon_{s_i}^d}\right)}$$
(refer to~\eqref{definition overline epsilon})
in $O(k^2)$ time and space.

Let $\mathcal{T}_k(n)$ be the time needed to compute $E_k^2$ on a set of $n$ points,
given the numbers $r_i$ and $s_i$
($0 \leq i \leq k$).
Following the proof of Theorem 2.1 in~\cite{nabilsaurabh},
we can compute $E_k^2$ by computing the following.
\begin{enumerate}
\item Let $\mathcal{H}$ be the set of all half-planes that contain at least $\overline{\epsilon_i^2}$ points of $P$. 
For each pair of half-planes $H,H'\in\mathcal{H}$,
there is a point $p_{H,H'}$ with minimum $y$-coordinate.
Compute the point with maximum $y$-coordinate
over all $p_{H,H'}$'s.
Langerman et al.~\cite{langerman2012computing} showed
how to compute that ``maximin'' point in $O(n\log^4(n))$ time.
\item Compute $E_{r_k}^2$ on a set of at most $n$ points,
which can be done in at most
$\mathcal{T}_{r_k}(n)$ time.
\item Compute $E_{r_s}^2$ on two different sets of at most $\overline{\epsilon_k^2}n$ points each,
which can be done in at most
$2\mathcal{T}_{s_k}\left(\overline{\epsilon_k^2}n\right)$ time.
\end{enumerate}
This leads to the following recurrence inequality.
\begin{align}
\label{ineq recurrence}
\mathcal{T}_k(n) \leq \mathcal{T}_{r_k}(n)+2\mathcal{T}_{s_k}\left(\overline{\epsilon_k^2}n\right) + O(n\log^4(n))
\end{align}
Using the facts that $k = 1+r_k+2s_k$ and $\overline{\epsilon_i^2} \leq 1$ for all $0\leq i \leq k$,
we can show that~\eqref{ineq recurrence} solves to $\mathcal{T}_k(n) = O(kn\log^4(n))$ by substitution.
Therefore,
the total time of computation is
$$O(k^2 + kn\log^4(n)) = O(kn\log^4(n))$$
since $k \leq n$.
\qed

\begin{figure}
\centering
\includegraphics[scale = 0.75]{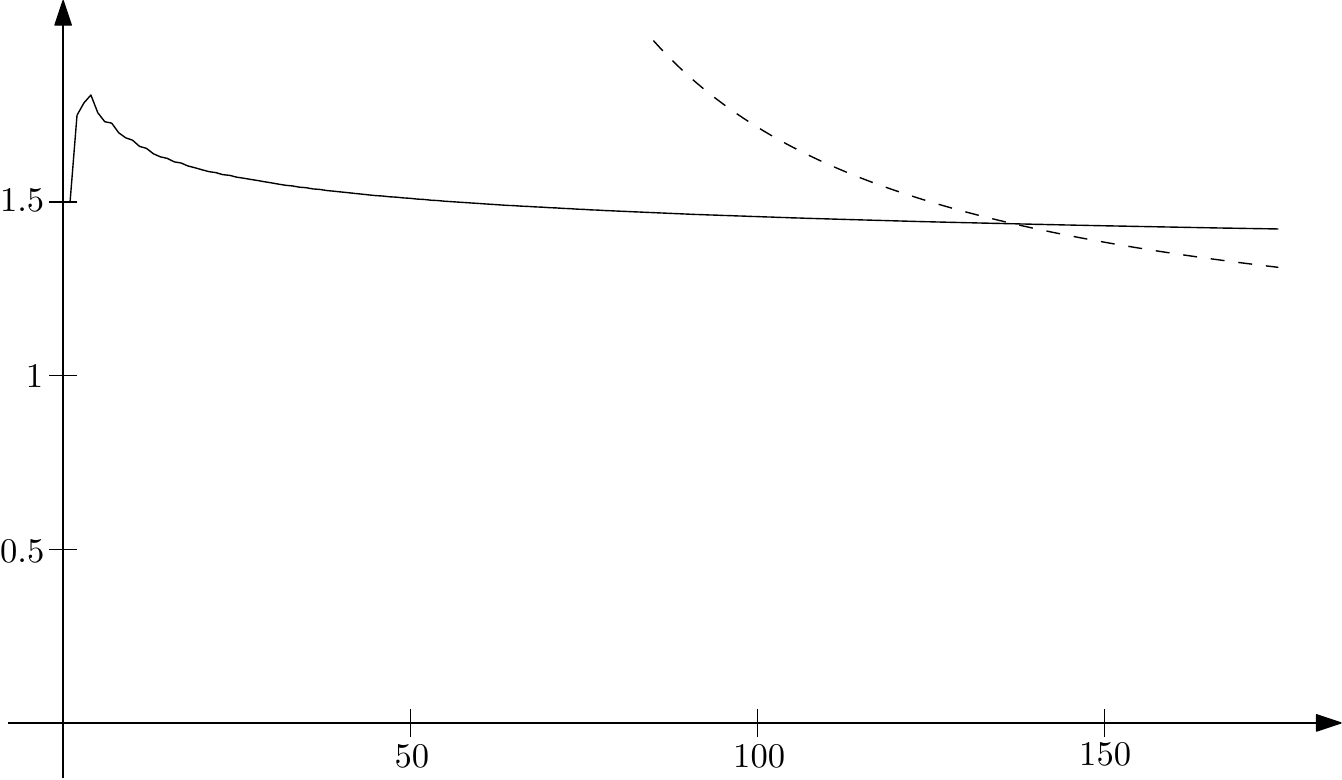}
\caption{Values of $\frac{2k-1}{2k\left(1-\overline{\epsilon_k^2}\right)}$ for $1\leq k \leq 175$ (plain curve). Values of $\frac{2k-1}{2(k-42)}$ for $85\leq k \leq 175$ (dashed curve).
\label{fig:compareFactors}}
\end{figure}

The values of $\frac{2k-1}{2k\left(1-\overline{\epsilon_k^2}\right)}$
are presented in Table~\ref{table 2D}
\begin{table}
\centering
\begin{tabular}{c|cccccccccc}
$k$ & $1$ & $2$ & $3$ & $4$ & $5$ & $6$ & $7$ & $8$ & $9$ & $10$ \\\hline
$\frac{2k-1}{2k\left(1-\overline{\epsilon_k^2}\right)}$ & $\frac{3}{2}$ & $\frac{7}{4}$ & $\frac{25}{14}$ & $\frac{217}{120}$ & $\frac{123}{70}$ & $\frac{187}{108}$ & $\frac{2249}{1302}$ & $\frac{1115}{656}$ & $\frac{91}{54}$ & $\frac{9633}{5740}$ \\
 & $1.50$ & $1.75$ & $1.79$ & $1.81$ & $1.76$ & $1.73$ & $1.73$ & $1.70$ & $1.69$ & $1.68$ \\
\end{tabular}
\caption{Values of $\frac{2k-1}{2k\left(1-\overline{\epsilon_k^2}\right)}$ for $1\leq k \leq 10$ together with numerical values.\label{table 2D}}
\end{table}
for $1\leq k \leq 10$
and in Figure~\ref{fig:compareFactors}
for $1\leq k \leq 175$.
Even though we do not know the optimal strategy for $\pone$,
we can guarantee that it wins when $k\geq 5$.
\begin{corollary}
\label{corollary pone wins 2D}
If $k \geq 5$,
then $\pone$ has a winning strategy for $VG(k,1)$.
\end{corollary}

\proof
We have
\begin{align*}
\text{payoff of $\pone$} &\geq \left(1-\overline{\epsilon_k^2}\right)n \\
& > \frac{1}{2} n,
\end{align*}
which can be verified numerically for $k \geq 5$
using the fact that
the sequence $\left(\overline{\epsilon_i^d}\right)_{i\geq 0}$ is decreasing.
\qed

In Subsection~\ref{section k >= 137},
we provide different bounds on the payoff of $\pone$
(refer to~\eqref{bounds pone 2D})
using a different approach.
The values of the approximation factors
from Subsection~\ref{section k >= 137}
are depicted in Figure~\ref{fig:compareFactors}
for $85\leq k \leq 175$.
We can prove that the strategy of the current subsection is better than the one of Subsection~\ref{section k >= 137} for $1\leq k \leq 136$
(refer to Proposition~\ref{proposition approach 1 better}).

\subsection{Bounds for $VG(k,1)$, Where $k \geq 137$}
\label{section k >= 137}

In this section,
we obtain a different approximation solution for $\pone$
by focusing on the reverse problem.
That is,
given any $0<\alpha<1$,
determine whether there exists an integer $\nu(\alpha)$ such that, in $VG(\nu(\alpha),1)$,
there is a placement strategy by $\pone$ such that $\ptwo$ can get at most $\alpha n$ users.
It is known that for any convex range space and any $0<\epsilon<1$,
there exists an $\epsilon$-net of size $O\left(\frac{1}{\epsilon}\textrm{polylog}\frac{1}{\epsilon}\right)$~\cite{chzell1995}.
Hence,
for any real number $0<\alpha<1$,
$\nu(\alpha)\in O\left(\frac{1}{\alpha}\textrm{polylog}\frac{1}{\alpha}\right)$.
The question is whether $\nu(\alpha)\in O(\frac{1}{\alpha})$?

We start with a technical lemma.
\begin{lemma}
\label{lemma technical alpha n/6}
Let $F_1$ be the set of facilities placed by $\pone$ and $f'$ be the facility placed by $\ptwo$, such that $f'$ serves at least $\alpha n$ users. There exists a circle which does not contain any of the facilities from $F_1$ and contains at least $\lceil\frac{\alpha n}{6}\rceil$ users.\end{lemma}

\begin{figure}
\centering
\includegraphics[scale = 0.625]{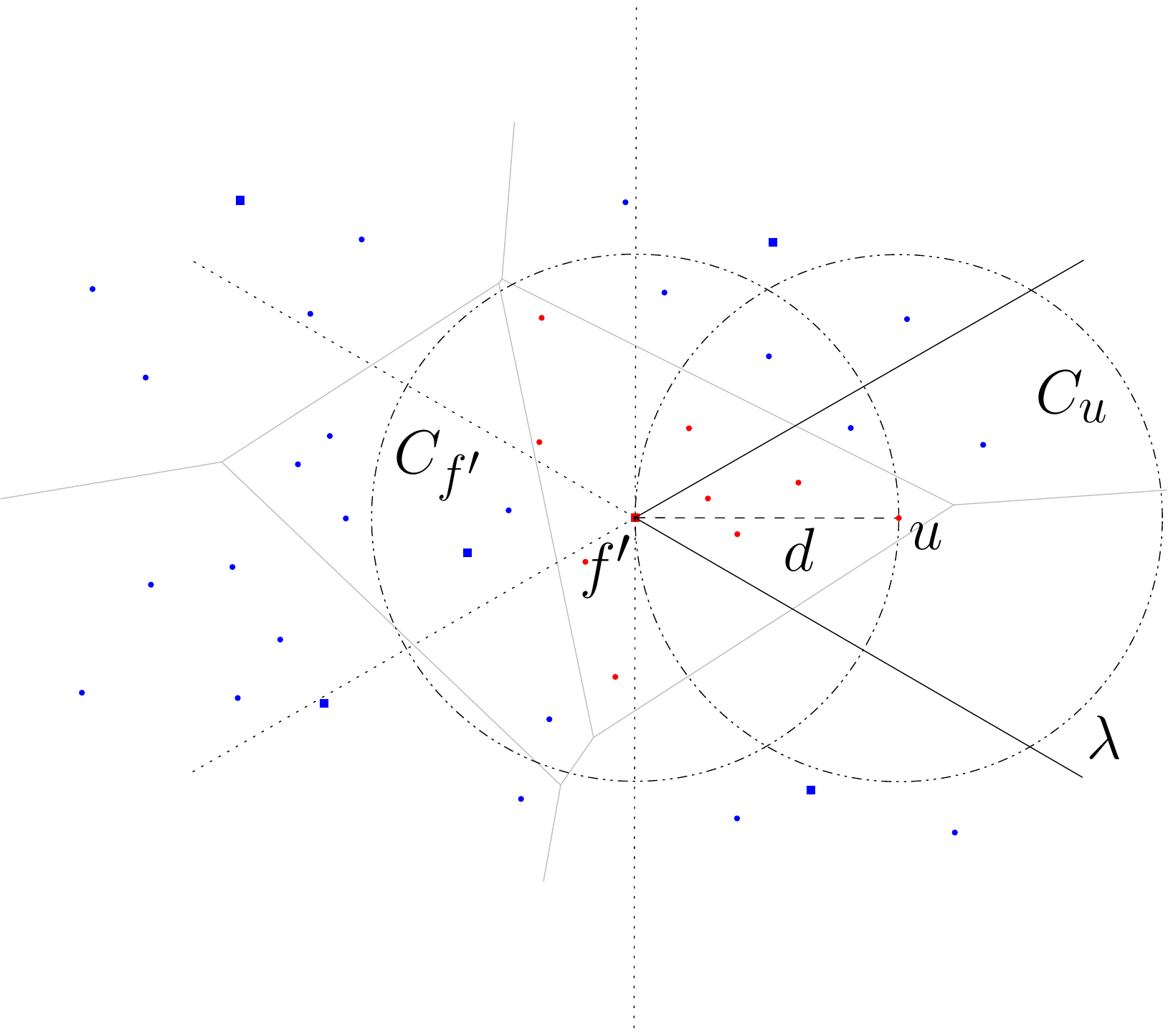}
\caption{Illustration of the proof of Lemma~\ref{lemma technical alpha n/6}.
The Voronoi diagram of $F_1 \cup \{f'\}$ is depicted in grey.
The points in red represent the $\alpha n$ users served by $\ptwo$,
i.e. by $f'$.\label{fig:divisionineqangle}}
\end{figure}

\proof
Denote by $U_{f'}$ the set of at least $\alpha n$ users served by $f'$.
Consider any six rays emerging from $f'$ such that
the angle between any two consecutive rays is $\frac{1}{3}\pi$
(refer to Figure~\ref{fig:divisionineqangle}).
These six rays divide the plane into six regions. At least one of these regions contains at least $\frac{\alpha n}{6}$ users from $U_{f'}$. Let $\lambda$ be such a region and $U_{\lambda} = U_{f'} \cap \lambda$ be the set of users in $\lambda$ that are served by $f'$.

Consider any user $u\in U_{\lambda}$ which is farthest from $f'$.
Denote the circle centered at $f'$ and passing through $u$ by $C_{f'}$.
Let the distance between $u$ and $f'$ be $d$.
Since $u\in U_{\lambda}$ is farthest from $f'$,
all the users in $\lambda$ that are served by $f'$
lie in the region $\lambda\cap C_{f'}$.
Since the angle between the bounding lines of $\lambda$ is $\frac{1}{3}\pi$,
the maximum distance between any two points in $\lambda\cap C_{f'}$ is $d$.
Hence,
the circle $C_u$ centered at $u$ with radius $d$ contains all the users in $\lambda\cap C_{f'}$ that are served by $f'$. Since $u$ is served by $f'$, $C_u$ does not contain any facility from $F_1$. Hence,
the result holds.
\qed

The following theorem is due to Matou\v{s}ek et al.~\cite{alot}.
\begin{theorem}[Matou\v{s}ek, Seidel and Welzl (1990)]
\label{thm matousek seidel welzl}
Let $0 <\epsilon\leq 1$ be a real number and let $\mathcal{D}$ be a family of disks.
For every finite point set $S$ in the plane there exists an $\epsilon$-net with respect to $\mathcal{D}$ of size $O(1/\epsilon)$.\label{th:epsilondisk}
\end{theorem}

We prove the following proposition using Lemma~\ref{lemma technical alpha n/6} and Theorem~\ref{th:epsilondisk}.
\begin{proposition}
\label{proposition loose bound}
For any real number $0<\alpha<1$, there exists an integer $k\in O(\frac{1}{\alpha})$ such that in $VG(k,1)$, $\pone$ can choose $k$ points to place its facilities such that the payoff of $\ptwo$ is at most $\alpha n$.
\end{proposition}

\proof
Fix $\epsilon=\frac{\alpha}{6}$.
Find an $\epsilon$-net $E$
such that any disk which contains $\epsilon n$ users contains at least one point from $E$.
By Theorem~\ref{thm matousek seidel welzl},
the size of $E$ is $k\in O(1/\epsilon) = O(1/\alpha)$.
We claim that if $\pone$ places its facilities at all points of $E$,
$\ptwo$ will get at most $\alpha n$ users.
Suppose there exists a placement of facility by $\ptwo$ which serves $\alpha n +1$ users.
From Lemma~\ref{lemma technical alpha n/6},
we know that there exists a disk which does not contain any point from $E$ and contains $\lceil\frac{\alpha n +1}{6}\rceil>\epsilon n$ users. This contradicts the fact that $E$ is an $\epsilon$-net.
\qed

From Proposition~\ref{proposition loose bound},
we know that $\nu(\alpha) \in O(1/\alpha)$.
However,
the constant hidden in the asymptotic notation is fairly big
(refer to~\cite{alot}).
Our next objective is to find a constant $\kappa$ as small as possible
such that $\nu(\alpha) = \kappa / \alpha$.

Let $0<\epsilon <1$ be any real number.
We know by Theorem~\ref{thm matousek seidel welzl} that for any $0 <\epsilon\leq 1$, there exists a set of $k\in O(1/\epsilon)$ points which pierces all the disks that contain $\epsilon n$ points. 
Next,
we prove that given any set $P$ of $n$ points, there exists a set of $7/\epsilon$ points which pierces any disk that contains $\epsilon n$ points.

Given a set $P$ of $n$ points, let the minimum disk that contains $\epsilon n$ points from $P$ be $D^*$. Consider the set $\mathcal{D}$ of all disks $D$ such that $D^*\cap D\neq {\O}$ and $D$ contains at least $\epsilon n$ points from $P$.
\begin{lemma}
\label{lemma:7isenough}
The set $\mathcal{D}$  can be pierced by $7$ points.
\end{lemma}

\begin{figure}
\centering
\includegraphics[scale = 1.125]{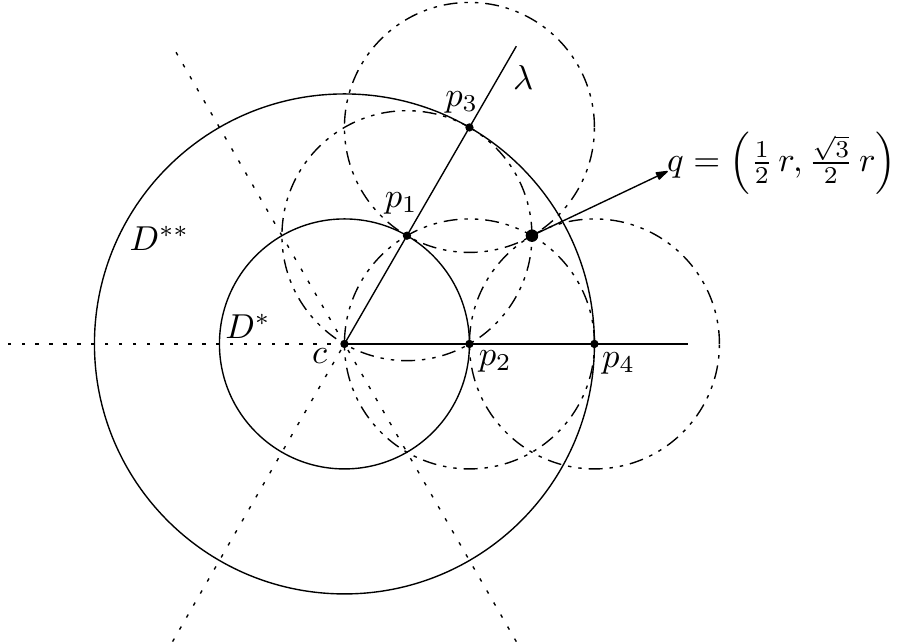}
\caption{Illustration of the proof of Lemma~\ref{lemma:7isenough}.\label{fig:mindiskb}}
\end{figure}

\proof
Let $D^*$ be centered at $c$ and denote the radius of $D^*$ by $r$.
Consider the disk $D^{**}$ centered at $c$ with radius $2r$.
We construct a set $Q$, which contains $7$ points,
that pierces $\mathcal{D}$.
Consider any six rays emerging from $c$ such that the angle between any two consecutive rays is $\frac{1}{3}\pi$
(refer to Figure~\ref{fig:mindiskb}).
These six rays define six sectors.
The set $Q$ consists in one point per sector, plus $c$.

Let $\lambda$ be any of the six sectors defined by the six rays.
Consider the set $\mathcal{D}_{\lambda}\subset \mathcal{D}$ of disks that do not contain $c$ and whose centers are in $\lambda$. We show that there exists a point that pierces $\mathcal{D}_{\lambda}$. 

Observe that the center of any disk in $\mathcal{D}_{\lambda}$  must lie outside of $D^*$ because $D^*$ is the minimum-radius disk that contains $\epsilon n$ points. The disks $D^*$ and $D^{**}$ intersect the boundary of $\lambda$ in four points
$p_1$, $p_2$, $p_3$ and $p_4$,
respectively (refer to Figure~\ref{fig:mindiskb}).
Without loss of generality,
suppose that $c=(0,0)$, $p_1=\left(\frac{1}{2}r,\frac{\sqrt{3}}{2}\,r\right)$, $p_2=(r,0)$ and $p_4 = (2r,0)$.
One can verify that $q=\left(\frac{3}{2}r,\frac{\sqrt{3}}{2} \,r\right)$
is such that $|p_1 q|=|p_2 q|=|p_3 q|=|p_4 q|=r$.
We prove that $q$ pierces $\mathcal{D}_{\lambda}$.
Let $c_i\in \lambda\setminus D^*$ be the center of a disk $D_i$ in  $\mathcal{D}_{\lambda}$.
We consider two cases:
(1) $c_i\in \lambda\setminus D^{**}$
or (2) $c_i\in \lambda\cap D^{**}$.
\begin{enumerate}
\item Without loss of generality,
assume that $c_i$ is on the line joining $p_2$ and $p_4$
(refer to Figure~\ref{fig:mindiskproofa}).
\begin{figure}
\centering
\includegraphics[scale = 1.125]{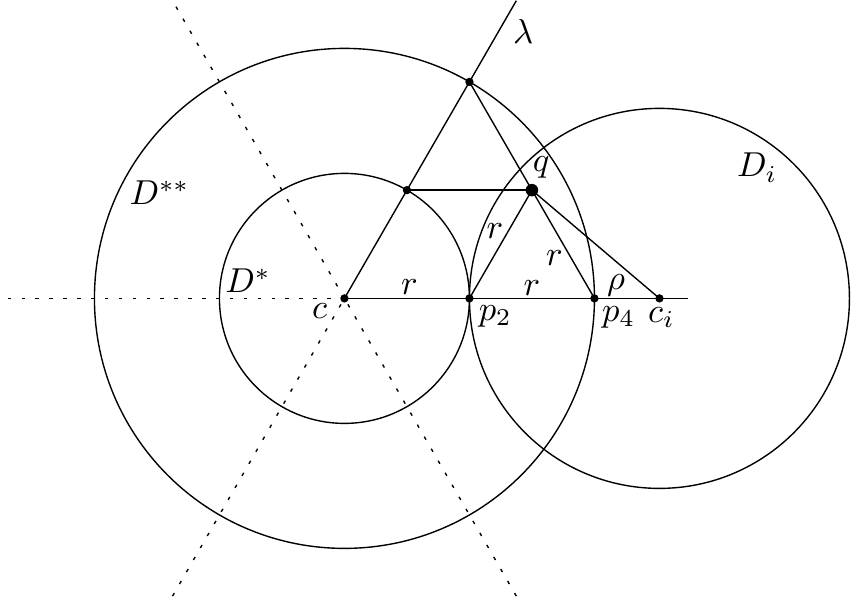}
\caption{Illustration of the proof of Lemma~\ref{lemma:7isenough}, Case (1).\label{fig:mindiskproofa}}
\end{figure}
From the definition of $q$,
we have $|qp_2| = |qp_4| = r$.
Furthermore,
from the definition of $D^{**}$,
we have $|p_2p_4| = r$.
Therefore,
the triangle $\triangle qp_2p_4$ is equilateral.
Hence,
the angle $\angle q p_2 p_4= \frac{1}{3}\pi$.
Let $\rho = |p_4c_i|$
and denote the radius of $D_i$ by $r_i$.
Since $D_i\cap D^*\neq {\O}$,
we have
\begin{equation} \label{eq:lemma3ia}
r_i\geq r+\rho.
\end{equation}
Consider the triangle $\triangle qp_4 c_i$.
From the triangle inequality,
we have
\begin{equation} \label{eq:lemma3ib}
r+\rho\geq |c_iq|.
\end{equation}
From~\eqref{eq:lemma3ia} and~\eqref{eq:lemma3ib},
we have $r_i\geq |c_iq|$.
Hence,
$q \in D_i$.

\item Without loss of generality,
assume that $c_i$ is on the line joining $p_2$ and $p_4$
(refer to Figure~\ref{fig:mindiskproofb}).
\begin{figure}
\centering
\includegraphics[scale=1.125]{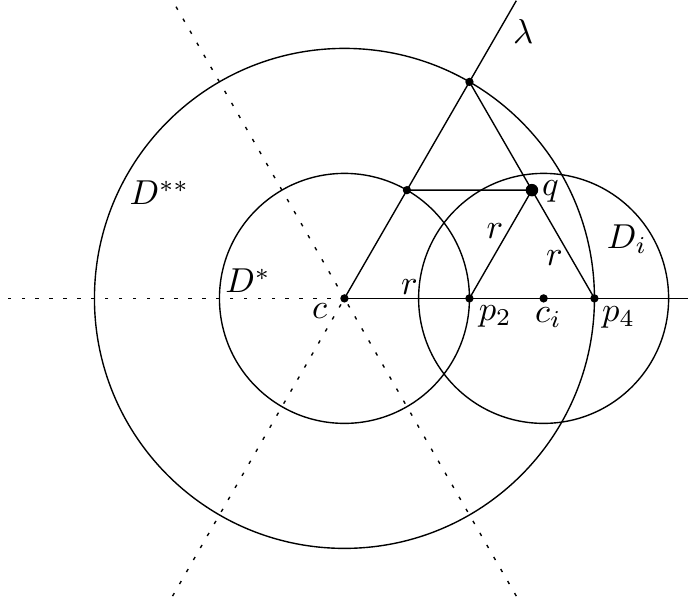}
\caption{Illustration of the proof of Lemma~\ref{lemma:7isenough}, Case (2).\label{fig:mindiskproofb}}
\end{figure}
Denote the radius of $D_i$ by $r_i$.
Since $D^*$ is the minimum-radius disk that contains $\epsilon n$ points,
we have $r_i \geq r$.
Since the triangle $\triangle q p_2 p_4$ is equilateral,
the distance from $q$ to any point on $p_2 p_4$ is less than $|p_2q|=r$.
Hence,
$q \in D_i$.
\end{enumerate}

Since $q \in D_i$ for all $D_i \in \mathcal{D}_{\lambda}$,
$q$ pierces $\mathcal{D}_{\lambda}$.
Add $q$ to $Q$ and repeat the same argument for each sector around $c$.
The set $Q$ pierces $\mathcal{D}$.
\qed

From Lemma~\ref{lemma:7isenough},
we can prove that there exists a weak $\epsilon$-net of size $7/\epsilon$.
\begin{theorem}
\label{th:epsilon7}
Given a set $P$ of $n$ points,
there exists a weak $\epsilon$-net $Q_{\epsilon}$ of size $7/\epsilon$ such that
any disk which contains $\epsilon n$ points from $P$ contains at least one point from $Q_{\epsilon}$.
\end{theorem}

\proof
We provide an iterative algorithm to construct $Q_{\epsilon}$.
At each stage of the algorithm,
we find the minimum disk $D^*$ that contains $\epsilon n$ points from $P$.
From Lemma~\ref{lemma:7isenough},
we know there exists a set $Q$ of $7$ points which pierces the set of all disks containing $\epsilon n$ points from $P$ and having a nonempty intersection with $D^*$.
We include these $7$ points in $Q_{\epsilon}$.
Then,
we remove all points of $P$ which are inside $D^*$.

We continue this process until $P$ contains no more than $\epsilon n$ points.
The cardinality of $Q_{\epsilon}$ at the end of the process is at most $7/\epsilon$.
We claim that this algorithm constructs a weak $\epsilon$-net.

Suppose $Q_{\epsilon}$ is not a weak $\epsilon$-net.
Denote by $D_i$ the minimum disk,
which contains $\epsilon n$ points from $P$,
that we choose at the $i$-th stage of the algorithm.
Since $Q_{\epsilon}$ is not a weak $\epsilon$-net,
there exists a disk $\hat{D}$,
which contains at least $\epsilon n$ points,
that is not pierced by any of the points in $Q_{\epsilon}$.
If $\hat{D}$ did not intersect any of the $D_i$'s,
then $\hat{D}$ would contain less than $\epsilon n$ points.
Therefore,
let $D_j$ be the first disk that has a nonempty intersection with $\hat{D}$.
Notice that none of the points in $\hat{D}$ have been removed from $P$ at earlier stages. Thus,
from Lemma~\ref{lemma:7isenough},
$\hat{D}$ must be pierced by one of the $7$ points chosen at stage $j$.
\qed

From Theorem~\ref{th:epsilon7} and Lemma~\ref{lemma technical alpha n/6},
we can prove the following theorem.
\begin{theorem}
For any real number $0<\alpha<1$,
there exists a placement of $\frac{42}{\alpha}$ facilities by $\pone$
such that $\ptwo$ can serve at most $\alpha n$ users by placing one facility. 
\end{theorem}

Thus,
from Lemma~\ref{lemma general upper bound},
\begin{align}
\label{bounds pone 2D}
\frac{k-42}{k}n \leq \text{payoff of $\pone$} \leq \frac{2k-1}{2k}n ,
\end{align}
provided that $k > 42$,
which leads to the following theorem.
\begin{theorem}
\label{theorem approx general k 7 points}
We can compute a $\frac{2k-1}{2(k-42)}$-factor approximation for $\pone$,
provided that $k > 42$.
\end{theorem}

The comparison of the approximation factor of Theorem~\ref{theorem approx general k}
with the approximation factor of Theorem~\ref{theorem approx general k 7 points}
is depicted in Figure~\ref{fig:compareFactors}.
We also have the following inequality.
\begin{proposition}
\label{proposition approach 1 better}
If $k \leq 136$,
$$\frac{2k-1}{2k\left(1-\epsilon_k^2\right)} < \frac{2k-1}{2(k-42)} .$$
\end{proposition}

\proof
\begin{align*}
\frac{2k-1}{2k\left(1-\epsilon_k^2\right)} &\leq \frac{2k-1}{2k\left(1-\overline{\epsilon_k^2}\right)} & \text{refer to the discussion of Section~\ref{section k <= 136},}\\
&< \frac{2k-1}{2(k-42)} & \text{can be verified numerically
(see Figure~\ref{fig:compareFactors}).}
\end{align*}
\qed

The running time of the algorithm from the proof of Theorem~\ref{th:epsilon7}
is related to the computation of the \emph{minimum $k$-enclosing disk}.
That is,
given a set $P$ of $n$ points and an integer $1\leq k \leq n$,
find the minimum disk that contains at least $k$ points from $P$.
In Table~\ref{table k-enclosing},
\begin{table}
\centering
\begin{tabular}{lccc}
Type of Algorithm & Space & Time of Computation & Reference \\\hline
Deterministic & $O(nk)$ & $O\left(nk\log^2(n)\right)$ & \cite{DBLP:journals/comgeo/EfratSZ94} \\
Deterministic & $O\left(n\log(n)\right)$ & $O\left(nk\log^2(n)\log(n/k)\right)$ & \cite{DBLP:journals/comgeo/EfratSZ94} \\
Deterministic & $O\left(n+k^2\right)$ & $O\left(n\log(n)+nk\right)$ & \cite{DBLP:journals/jal/DattaLSS95} \\
Randomized & $O\left(n+k^2\right)$ & $O(nk)$ & \cite{DBLP:journals/algorithmica/Har-PeledM05} \\
\end{tabular}
\caption{Best known algorithms to compute the minimum $k$-enclosing disk of a set of $n$ points.\label{table k-enclosing}}
\end{table}
we present the best known algorithms to solve this problem.

In the algorithm from the proof of Theorem~\ref{th:epsilon7},
we have $k = \epsilon n$
and we need to compute the minimum $k$-enclosing disk $1/\epsilon$ times.
Therefore,
from the algorithms of Table~\ref{table k-enclosing},
we directly get algorithms to construct the weak $\epsilon$-net
of Theorem~\ref{th:epsilon7}.
The running times of these algorithms
are presented in Table~\ref{table time computation 7 epsilon}.
\begin{table}
\centering
\begin{tabular}{lcc}
Type of Algorithm & Storage & Time of Computation \\\hline
Deterministic & $O\left(\epsilon n^2\right)$ & $O\left(n^2\log^2(n)\right)$ \\
Deterministic & $O\left(n\log(n)\right)$ & $O\left(n^2\log^2(n)\log(1/\epsilon)\right)$ \\
Deterministic & $O\left(n+\epsilon^2n^2\right)$ & $O\left(\frac{1}{\epsilon}n\log(n)+n^2\right)$ \\
Randomized & $O\left(n+\epsilon^2n^2\right)$ & $O\left(n^2\right)$ \\
\end{tabular}
\caption{Time of computation for a weak $\epsilon$-net of size $7/\epsilon$.\label{table time computation 7 epsilon}}
\end{table}

\section{Voronoi Game in $3$ Dimensions}
\label{section 3D}

In this section,
we study $VG(k,1)$ in three dimensions.
We translate the results and the proofs from Section~\ref{section 2D} into three dimensions.

\subsection{$VG(k,1)$ in $3$ Dimensions for $k \leq 805$}

This subsection
is the three-dimensional version of Subsection~\ref{section k <= 136}.
We start with the following lemma.
\begin{lemma}
\label{lemma general upper bound 3D}
Let $\{f_1,f_2,...,f_k\} \subset \mathbb{R}^3$ be the set of locations of the $k$ facilities placed by $\pone$.
There is a strategy for $\ptwo$ that guarantees a payoff of at least $\frac{1}{2k}n$ for $\ptwo$.
\end{lemma}

\proof
The proof is identical to that of Lemma~\ref{lemma general upper bound}
by considering half-spaces instead of half-planes.
\qed

Then,
using Corollary~\ref{cor:rayndmustafa general},
we can prove the following lemma.
\begin{lemma}
If $\pone$ places its facilities at all points of $E_k^3$,
respectively,
then the payoff of $\ptwo$ is at most $\overline{\epsilon_k^3}n$.
\end{lemma}

\proof
The proof is identical to that of Lemma~\ref{lemma ptwo at most epsilon k 2}
since three-dimensional Voronoi cells are convex.
\qed

\begin{figure}
\centering
\includegraphics[scale = 0.75]{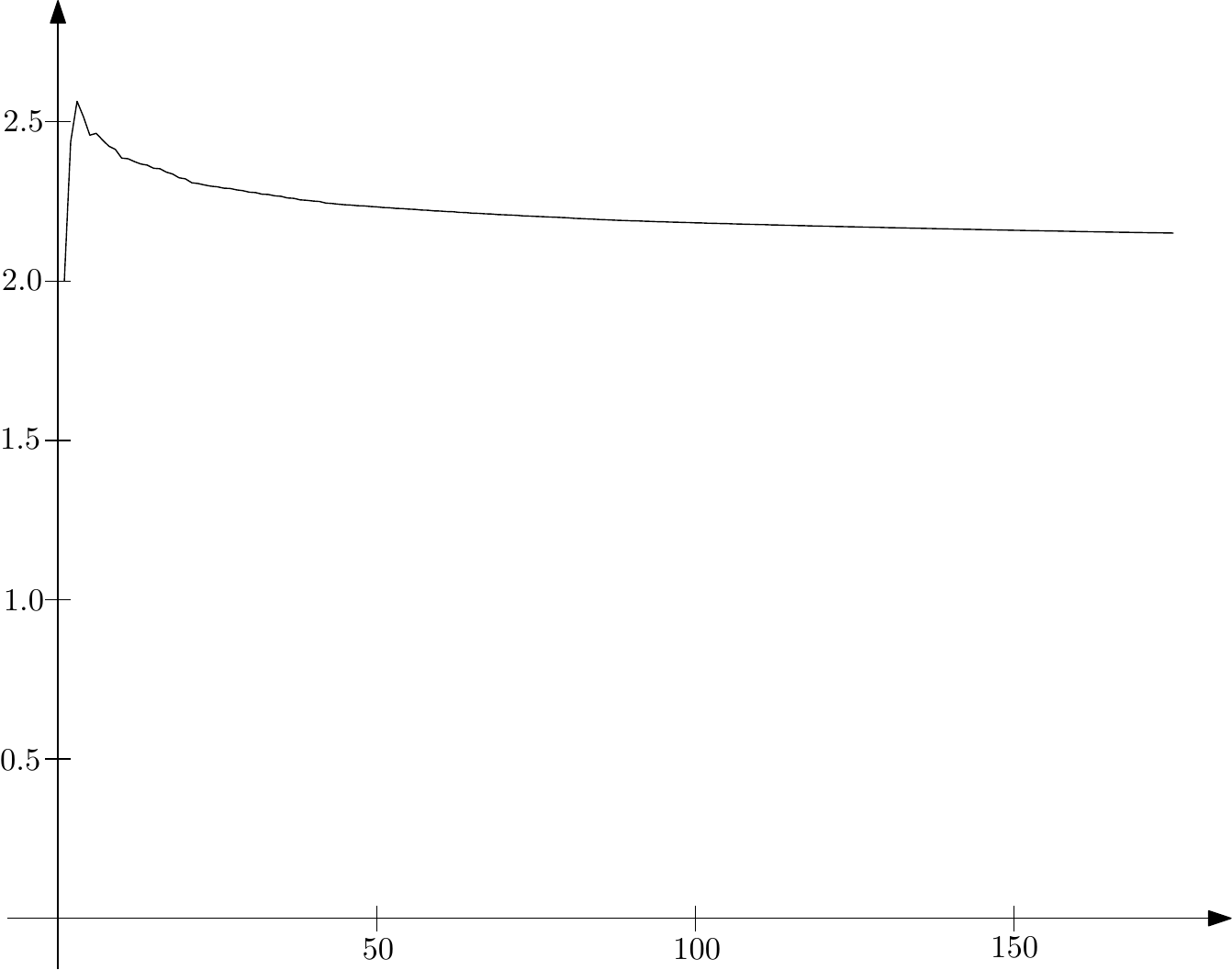}
\caption{Values of $\frac{2k-1}{2k\left(1-\overline{\epsilon_k^3}\right)}$ for $1\leq k \leq 175$.
\label{fig:compareFactors-3D}}
\end{figure}

\begin{figure}
\centering
\includegraphics[scale = 0.75]{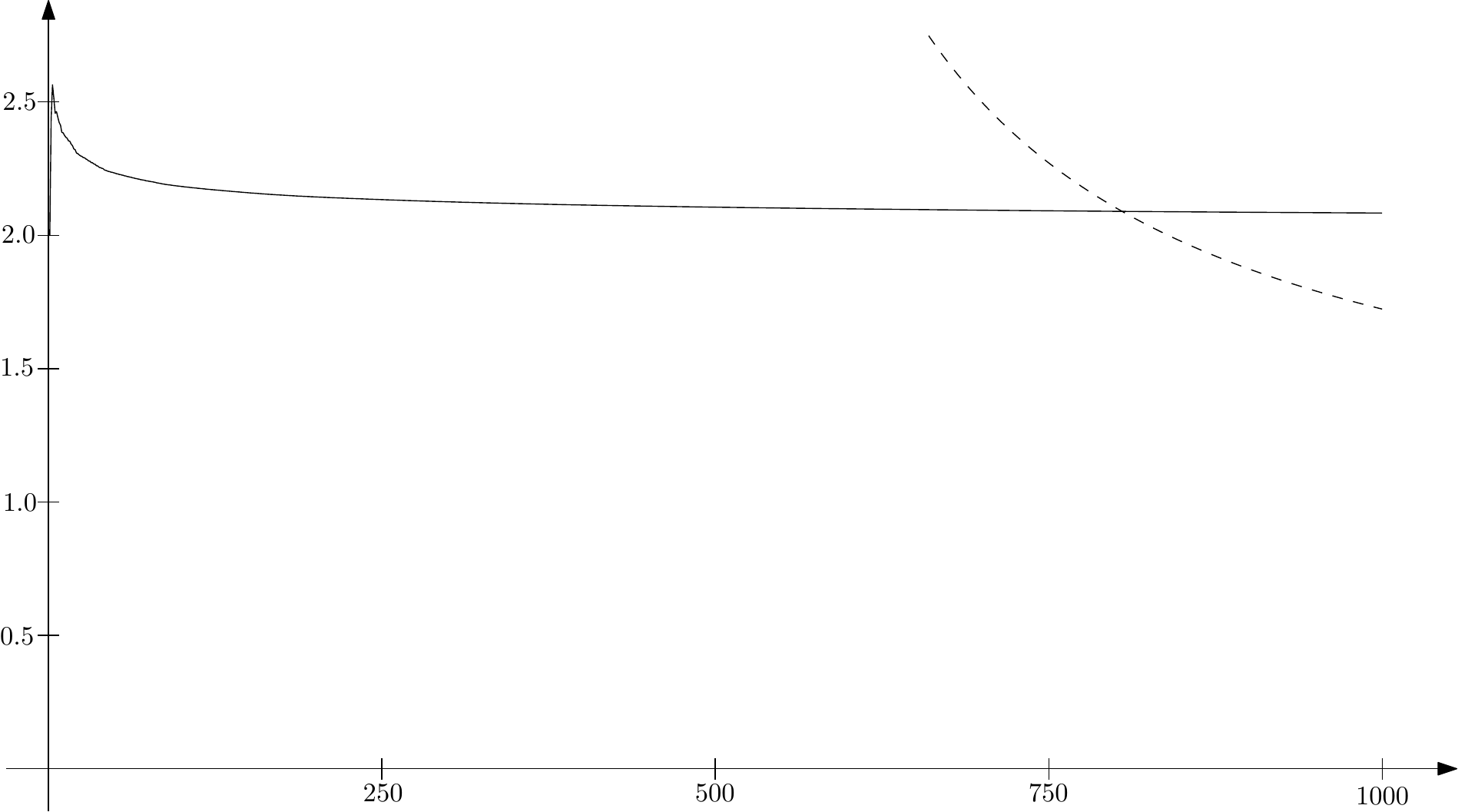}
\caption{Values of $\frac{2k-1}{2k\left(1-\overline{\epsilon_k^3}\right)}$ for $1\leq k \leq 1000$ (plain curve).
Values of $\frac{2k-1}{2(k-420)}$ for $660\leq k \leq 1000$ (dashed curve).
\label{fig:compareFactors-3Db}}
\end{figure}

Therefore,
$$\left(1-\overline{\epsilon_k^3}\right)n \leq \text{payoff of $\pone$} \leq \frac{2k-1}{2k}n .$$
Consequently,
we get a $\frac{(2k-1)n/2k}{(1-\overline{\epsilon_k^3})n} = \frac{2k-1}{2k(1-\overline{\epsilon_k^3})}$-factor approximation of the optimal strategy for $\pone$.
The values of $\frac{2k-1}{2k\left(1-\overline{\epsilon_k^3}\right)}$
are presented in Table~\ref{table 3D}
\begin{table}
\centering
\begin{tabular}{c|cccccccccc}
$k$ & $1$ & $2$ & $3$ & $4$ & $5$ & $6$ & $7$ & $8$ & $9$ & $10$ \\\hline
 & $2$ & $\frac{39}{16}$ & $\frac{100}{39}$ & $\frac{161}{64}$ & $\frac{639}{260}$ & $\frac{473}{192}$ & $\frac{1573}{644}$ & $\frac{5505}{2272}$ & $\frac{6494}{2691}$ & $\frac{22021}{9230}$ \\ 
 & $2.00$ & $2.44$ & $2.56$ & $2.52$ & $2.46$ & $2.46$ & $2.44$ & $2.42$ & $2.41$ & $2.39$
\end{tabular}
\caption{Values of $\frac{2k-1}{2k\left(1-\overline{\epsilon_k^3}\right)}$ for $1\leq k \leq 10$ together with numerical values.\label{table 3D}}
\end{table}
for $1\leq k \leq 10$,
in Figure~\ref{fig:compareFactors-3D}
for $1\leq k \leq 175$
and in Figure~\ref{fig:compareFactors-3Db})
for $1\leq k \leq 1000$.

In Section~\ref{section k >= 806},
we provide different bounds on the payoff of $\pone$
(refer to~\eqref{bounds pone 3D})
using a different approach.
The values of the approximation factors
from Subsection~\ref{section k >= 806}
is depicted in Figure~\ref{fig:compareFactors-3Db}
for $660\leq k \leq 1000$.
We can prove that the strategy of the current subsection is better than the one of Subsection~\ref{section k >= 806} for $1\leq k \leq 805$
(refer to Proposition~\ref{proposition approach 1 better 3D}).

\subsection{$VG(k,1)$ in $3$ Dimensions for $k \geq 806$}
\label{section k >= 806}

Let $D$ be a disk in the plane with radius $r$ and center $c$.
Consider any six rays emerging from $c$ such that
the angle between any two consecutive rays is $\frac{1}{3}\pi$.
These six rays divide $D$ into six circular sectors.
These circular sectors satisfy certain properties that we used
in the proofs of Lemmas~\ref{lemma technical alpha n/6}
and~\ref{lemma:7isenough}.
In this section,
we translate these properties in $\mathbb{R}^3$
and obtain three dimensional versions of the results of Section~\ref{section k >= 137}.
We start with the following lemma.
\begin{lemma}
\label{lemma 20 cones}
Let $B$ be a ball in $\mathbb{R}^3$ with center $c$.
We can cover the whole volume of $B$
using $20$ cones with apex $c$ and aperture $\frac{1}{3}\pi$.
\end{lemma}

\proof
Tarnai and G{\'a}sp{\'a}r~\cite[Table 2]{tarnai1991covering} proved that
we can cover the surface of a ball
using $20$ circles with angular radius less than $0.165\pi < \frac{1}{6}\pi$.
Hence,
the surface of a ball
can be covered using $20$ circles with angular radius $\frac{1}{6}\pi$.
With each of these circles,
we can construct a cone with apex $c$.
Since the circles cover the surface of $B$,
the cones cover $B$.
Moreover,
since the angular radius of these circles is $\frac{1}{6}\pi$,
the aperture of the cones is $\frac{1}{3}\pi$.
\qed

Let $C$ be a cone with aperture $\frac{1}{3}\pi$ and radius $r$.
Any ball with radius $r$ and center in $C$ contains $C$.
Together with Lemma~\ref{lemma 20 cones},
this leads to the following lemma.
\begin{lemma}
\label{lemma technical alpha n/20}
Let $F_1$ be the set of facilities placed by $\pone$ and $f'$ be a facility placed by $\ptwo$,
such that $f'$ serves at least $\alpha n$ users.
There exists a ball which does not contain any of the facilities from $F_1$
and contains at least $\lceil\frac{\alpha n}{20}\rceil$ users.
\end{lemma}

\proof
Denote by $U_{f'}$ the set of at least $\alpha n$ users served by $f'$.
Let $B$ be the unit ball centered at $f'$.
Consider the $20$ cones from Lemma~\ref{lemma 20 cones} that cover $B$.
At least one of these cones contains at least $\frac{\alpha n}{20}$ users from $U_{f'}$.
Let $\lambda$ be such a region and $U_{\lambda} = U_{f'} \cap \lambda$
be the set of users in $\lambda$ that are served by $f'$.

Consider any user $u\in U_{\lambda}$ which is farthest from $f'$.
Denote the ball centered at $f'$ and passing through $u$ by $B_{f'}$.
Let the distance between $u$ and $f'$ be $d$.
Since $u\in U_{\lambda}$ is farthest from $f'$,
all the users in $\lambda$ that are served by $f'$
lie in the region $\lambda\cap B_{f'}$.
Since $\lambda$ has an aperture of $\frac{1}{3}\pi$,
the maximum distance between any two points in $\lambda\cap B_{f'}$ is $d$.
Hence,
the ball $B_u$ centered at $u$ with radius $d$ contains all the users in $\lambda\cap B_{f'}$ that are served by $f'$.
Since $u$ is served by $f'$,
$B_u$ does not contain any facility from $F_1$.
Hence,
the result holds.
\qed

Let $0<\epsilon <1$ be any real number.
We now explain how to construct a weak $\epsilon$-net $Q_{\epsilon }$
of size $\frac{420}{\epsilon}$ such that
any ball which contains $\epsilon n$ points from $P$
contains at least one point from $Q_{\epsilon}$.
Given a set $P$ of $n$ points,
let the minimum ball that contains $\epsilon n$ points from $P$ be $B^*$.
Consider the set $\mathcal{B}$ of all balls $B$
such that $B^*\cap B\neq {\O}$
and $B$ contains at least $\epsilon n$ points from $P$.
\begin{lemma}
\label{lemma:21isenough}
The set $\mathcal{B}$ can be pierced by $21$ points.
\end{lemma}

\proof
Let $B^*$ be centered at $c$ and denote the radius of $B^*$ by $r$.
Consider the ball $B^{**}$ centered at $c$ with radius $2r$.
We construct a set $Q$, which contains $21$ points,
that pierces $\mathcal{B}$.
Consider any of the $20$ cones from Lemma~\ref{lemma 20 cones} that cover $B^*$.
These $20$ cones define $20$ sectors.
The set $Q$ consists in one point per sector, plus $c$.
Let $\lambda$ be any of the $20$ sectors defined by the $20$ cones.
Consider the set $\mathcal{B}_{\lambda}\subset \mathcal{B}$ of balls
that do not contain $c$ and whose centers are in $\lambda$.
We show that there exists a point that pierces $\mathcal{B}_{\lambda}$. 

\begin{figure}
\centering
\includegraphics[scale = 1]{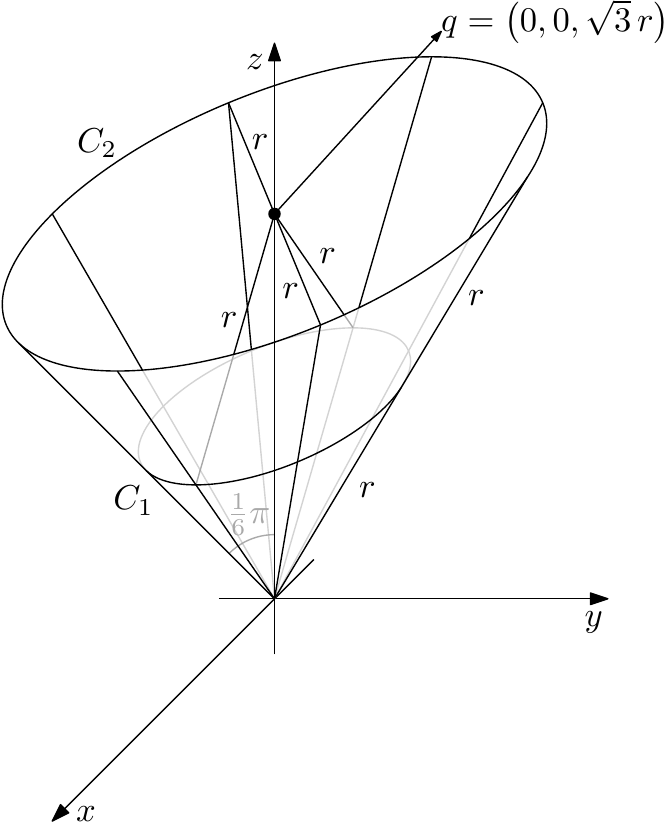}
\caption{Illustration of the proof of Lemma~\ref{lemma:21isenough}.\label{fig:cone3D}}
\end{figure}

Observe that the center of any ball in $\mathcal{B}_{\lambda}$ must lie outside of $B^*$ because $B^*$ is the minimum-radius ball that contains $\epsilon n$ points.
The disks $B^*$ and $B^{**}$ intersect the boundary of $\lambda$
in two circles $C_1$ and $C_2$,
respectively.
Without loss of generality,
suppose that the boundary of $\lambda$  satisfies the equation $z = \sqrt{3(x^2+ y^2)}$
(refer to Figure~\ref{fig:cone3D}).
Therefore,
$C_1$ satisfies $x^2+y^2=\frac{1}{4}\,r^2$ and $z = \frac{\sqrt{3}}{2}\,r$,
and $C_2$ satisfies $x^2+y^2=r^2$ and $z =\sqrt{3}\,r$.
One can verify that $q = \left(0,0,\sqrt{3}\,r\right)$ is such that
$|pq| = r$ for all $p\in C_1$ and all $p\in C_2$.

The rest of the proof is similar to that of Lemma~\ref{lemma:7isenough}.
\qed

From Lemma~\ref{lemma:21isenough},
we can prove that there exists a weak $\epsilon$-net of size $21/\epsilon$.
\begin{theorem}
\label{th:epsilon21}
Given a set $P$ of $n$ points,
there exists a weak $\epsilon$-net $Q_{\epsilon}$ of size $21/\epsilon$ such that
any ball which contains $\epsilon n$ points from $P$ contains at least one point from $Q_{\epsilon}$.
\end{theorem}

\proof
The proof is identical to that of Theorem~\ref{th:epsilon21}.
\qed

From Theorem~\ref{th:epsilon21} and Lemma~\ref{lemma technical alpha n/20},
we can prove the following theorem.
\begin{theorem}
For any real number $0<\alpha<1$,
there exists a placement of $\frac{420}{\alpha}$ facilities by $\pone$
such that $\ptwo$ can serve at most $\alpha n$ users by placing one facility. 
\end{theorem}

Thus,
from Lemma~\ref{lemma general upper bound 3D},
\begin{align}
\label{bounds pone 3D}
\frac{k-420}{k}n \leq \text{payoff of $\pone$} \leq \frac{2k-1}{2k}n ,
\end{align}
provided that $k > 420$,
which leads to the following theorem.
\begin{theorem}
\label{theorem approx general k 21 points}
We can compute a $\frac{2k-1}{2(k-420)}$-factor approximation for $\pone$,
provided that $k > 420$.
\end{theorem}

The comparison of the approximation factor of Theorem~\ref{theorem approx general k 3D}
with the approximation factor of Theorem~\ref{theorem approx general k 21 points}
is depicted in Figure~\ref{fig:compareFactors-3Db}.
We also have the following inequality.
\begin{proposition}
\label{proposition approach 1 better 3D}
If $k \leq 805$,
$$\frac{2k-1}{2k\left(1-\epsilon_k^3\right)} < \frac{2k-1}{2(k-420)} .$$
\end{proposition}

\proof
\begin{align*}
\frac{2k-1}{2k\left(1-\epsilon_k^3\right)} &\leq \frac{2k-1}{2k\left(1-\overline{\epsilon_k^3}\right)} & \text{refer to the discussion of Section~\ref{section k <= 136},}\\
&< \frac{2k-1}{2(k-420)} & \text{can be verified numerically
(see Figure~\ref{fig:compareFactors-3Db}).}
\end{align*}
\qed

The running time of the algorithm from the proof of Theorem~\ref{th:epsilon21}
is related to the computation of the minimum $k$-enclosing ball in three dimensions.
That is,
given a set $P$ of $n$ points and an integer $1\leq k \leq n$,
find the minimum disk that contains at least $k$ points from $P$.
There is an algorithm by Datta et al.~\cite{DBLP:journals/jal/DattaLSS95}
to solve that problem in $O\left(n\log(n)+nk^2\log^2(k)\right)$ time
using $O\left(n+k^3\log(k)\right)$ space.
In the algorithm from the proof of Theorem~\ref{th:epsilon21},
we have $k = \epsilon n$
and we need to compute the minimum $k$-enclosing ball $1/\epsilon$ times.
This leads directly to an algorithm to construct the weak $\epsilon$-net
of Theorem~\ref{th:epsilon21}.
The runing time is $O\left(\frac{1}{\epsilon}n\log(n)+\epsilon n^3\log^2(\epsilon n)\right)$
and it uses $O\left(n+\epsilon^3n^ 3\log(\epsilon n)\right)$ space.

Even though we do not know the optimal strategy for $\pone$,
we can guarantee that it wins when $k\geq 841$.
\begin{corollary}
If $k \geq 841$,
then $\pone$ has a winning strategy for $VG(k,1)$.
\end{corollary}

\proof
We have
\begin{align*}
\text{payoff of $\pone$} &\geq \frac{k-420}{k}n &\text{by~\eqref{bounds pone 3D},} \\
&> \frac{1}{2}n &\text{since $k \geq 841$.}
\end{align*}
\qed

\section{Conclusion}
\label{section conclusion}

As we explained in Subsection~\ref{subsection preliminaries},
finding an exact solution to $VG(k,1)$
is a challenging task.
By following existing techniques used to solve Voronoi games,
we would get polynomial time algorithms to find the optimal strategy for $\pone$,
where the polynomial has a very high degree.
This is why our goal in this paper was to find approximate solutions
with significantly better running times.
We studied $VG(k,1)$ for $k \geq 1$ in two and three dimensions.

In two dimensions,
we find two different strategies,
both of them based on weak $\epsilon$-nets.
The first one
(refer to Subsections~\ref{subsection label VG(1,1) 2D},
\ref{subsection label VG(2,1) 2D}
and~\ref{section k <= 136})
was based on weak $\epsilon$-nets for convex sets.
More precisely,
we used a result by Mustafa and Ray~\cite{nabilsaurabh}
(refer to Corollary~\ref{cor:rayndmustafa general}).
The running times and approximation factors we get from that strategy are presented in Table~\ref{table conclusion 2D}.
\begin{table}
\centering
\begin{tabular}{c|cc}
$k$ & Approximation Factor & Running time \\\hline
$1$ & $3/2$ & $O(n)$ \\
$2$ & $7/4$ & $O(n\log^4(n))$ \\
$3$ & $25/14$ & $O(n\log^4(n))$ \\
$4$ & $217/120$ & $O(n\log^4(n))$ \\
$5$ & $123/70$ & $O(n\log^4(n))$ \\
$k> 5$ & $\frac{2k-1}{2k(1-\epsilon_k^2)}$ & $O(kn\log^4(n))$ \\
\end{tabular}
\caption{Running times and approximation factors for $VG(k,1)$ in two dimensions,
where $1\leq k \leq 5$.\label{table conclusion 2D}}
\end{table}
With that approach,
we can proved that for $k\geq 5$,
$\pone$ can win $VG(k,1)$ with an approximate solution
(refer to Corollary~\ref{corollary pone wins 2D}).
The second strategy
(refer to Subsection~\ref{section k >= 137})
was based on weak $\epsilon$-nets for circles.
We first proved that there exists weak $\epsilon$-nets for circles of size $7/\epsilon$
(refer to Theorem~\ref{th:epsilon7})
and then used that result to approximate the payoff of $\pone$.
When $k > 42$,
we get a $\frac{2k-1}{2(k-42)}$-approximation solution
(refer to Theorem~\ref{theorem approx general k 7 points})
for which the running time is directly related to the solution of the minimum $k$-enclosing disk
(refer to Table~\ref{table k-enclosing}).
This leads to a polynomial time algorithm,
where the polynomial has a low degree.
For $VG(k,1)$ in three dimensions,
we get similar results (refer to Section~\ref{section 3D}).
However,
we do not know how to compute $E_k^3$ efficiently.
For the two-dimensional case,
we could describe a recursive algorithm to compute $E_k^2$
(refer to Theorem~\ref{theorem approx general k}).
For the three-dimensional case,
we leave this question as an open problem.


The game $VG(k,l)$,
both from the exact and approximate points of view,
leads to numerous open questions.
What is the optimal way of solving $VG(k,l)$ exactly
(for $l\geq 1$)?
What is the connection between $VG(k,l)$ and weak $\epsilon$-nets
(for $l > 1$)?
When $l = 1$,
do we have to compute the minimum $k$-enclosing disk
to build a weak $\epsilon$-net of size $7/\epsilon$
(or $21/\epsilon$ in three dimensions)?
If not,
can we improve the time of computation
of the minimum $k$-enclosing disk of a set of points?
Is there a way of constructing weak $\epsilon$-nets of size smaller than $7/\epsilon$
(or $21/\epsilon$ in three dimensions)?
Finally,
how tight are our bounds on the payoff of $\pone$?
For instance,
are there sets $U$ of users for which $\pone$ cannot get more than $\left(1-\epsilon_k^2\right)n$ users
(or $\left(1-\epsilon_k^3\right)n$ users in three dimensions)?
Are there sets $U$ of users for which $\pone$ can achieve a payoff of $\frac{2k-1}{2k}n$?
Chawla et al.~\cite{Chawla:2004:WPL:988772.988815,DBLP:journals/orl/ChawlaRRS06}
studied these questions in a different framework.
We need to investigate further to see whether their techniques can be applied on our framework.

\bibliographystyle{plain}
\bibliography{voronoi_game_ref}

\end{document}